\documentstyle[emulateapj]{article}

\lefthead{Mosquera Cuesta \& Salim}

\righthead{Nonlinear electrodynamics and pulsar surface redshift}

\newcommand{\be}{\begin{equation}}

\newcommand{\ee}{\end{equation}}

\begin{document}

\title{Nonlinear electrodynamics and the surface redshift of pulsars}

\author{Herman J. Mosquera Cuesta\altaffilmark{1,2,3} and Jos\'e M. Salim\altaffilmark{1} }

\altaffiltext{1}{Centro
Brasileiro de Pesquisas F\'{\i}sicas, Laborat\'orio de Cosmologia e
F\'{\i}sica Experimental de Altas Energias \\Rua Dr. Xavier Sigaud 150,
Cep 22290-180, Urca, Rio de Janeiro, RJ, Brazil ---- e-mails: 
hermanjc@cbpf.br ::: jsalim@cbpf.br }
\altaffiltext{2}{Abdus Salam International Centre for Theoretical Physics,
Strada Costiera 11, Miramare 34014, Trieste, Italy}
\altaffiltext{3}{Centro Latino-Americano de F\'{\i}sica, Avenida Wenceslau 
Braz 71, CEP 22290-140 Fundos, Botafogo, Rio de Janeiro, RJ, Brazil}



\begin{abstract}
Currently is argued that the best method of determining the neutron
star (NS) fundamental properties is by measuring the {\it gravitational
redshift} ($z$) of spectral lines produced in the star photosphere.
Measurement of $z$ at the star surface provides a unique insight on the
NS mass-to-radius relation and thus on its equation of state (EoS),
which reflects the physics of the strong interaction between particles
making up the star. Evidence for such a measurement has been provided
quite recently by Cottam, Paerels \& Mendez (2002), and also by Sanwal
et { al.} (2002). Here we argue that although the quoted
observations are undisputed for canonical pulsars, they could be
misidentified if the NS is endowed with a super strong $B$ as in the
so-called magnetars (Duncan \& Thompson 1992) and strange quark
magnetars (Zhang 2002), as in the spectral line discovered by Ibrahim
et { al.} (2002, 2003). The source of this new ``confusion" redshift
is related to nonlinear electrodynamics (NLED) effects.


\end{abstract}


\keywords{Gravitation: redshift  --- line: formation --- line: identification 
--- stars: magnetic fields --- nonlinear methods: electrodynamics
 --- stars: pulsars: general }

\def\be{\begin{equation}}

\def\ee{\end{equation}}




\section{INTRODUCTION}


Neutron stars (NSs), the death throes of massive stars,  are among the 
most exotic objects in the universe. They are supposed to be composed of
essentially neutrons, although some protons and electrons are also
required in order to guarantee stability against Pauli's exclusion
principle for fermions. In view of its density, a neutron star is also
believed to trap in its core a substantial part of even more exotic
states of matter (EoS). It is almost a concensus that these new states
might exist inside and may dominate the star structural properties.
Pions plus kaons Bose-Einstein condensates could appear, as well as
``bags" of strange quark matter (Miller 2002). This last one believed
to be the most stable state of nuclear matter (Glendenning 1997), which
implies an extremely dense medium whose physics is currently under
severe scrutiny.  The major effect of these exotic constituents is
manifested through the NS mass-radius ratio ($M/R$). Most researchers
in the field think of the presence of such exotic components not only
as to make the star more compact, i.e., smaller in radius, but also to
lower the maximum mass it can retain. To get some insight into the
neutron star most elusive properties: its mass ($M)$ and radius ($R$),
astronomers use several techniques at disposal, being the most 
prospective one the {\it gravitational redshift}. Since the redshift
depends on the ratio $M/R$, then measuring NS spectral lines
displacement leads to a direct insight into this dense matter 
equation of state.

In the late years strong evidence seems to have gathered around a new
and exotic class of hyper magnetized neutron stars: the so-called
``magnetars" (Duncan \& Thompson 1992).  These objects are supposed to
be the final stage of newly-born neutron stars in which a
classical alpha-omega dynamo mechanism has efficiently acted on during
the early stages of its evolution, reaching field strentghs up to
$B_{\rm Sup-Crit} \sim 10^{17}$~G. A peculiar class of gamma-ray
sources known as ``soft gamma-ray repeaters" (SGRs) (Kouveliotou et
{ al.} 1998), and a set of X-ray pulsars known as of ``anomalous"
(AXPs), have been claimed to be associated with these type of stars
(Mereghetti 1999).

Although these magnetars are said to be the best model to explain
the dynamics of SGRs and AXPs, accretion-driven models
(Marsden et { al.} 2001), strange quark matter stars with normal 
$B$s (Zhang, Xu \& Qiao 2000; Xu \& Busse 2001; Hu \& Xu 2002; Xu 
2002),  or even 
the strange magnetar interpretation (Zhang 2002) have also been 
proposed as competing scenarios. Note in passing that in a recent 
paper P\'erez Mart\'{\i}nez et { al.}  (2003) have provided
arguments contending the formation itself of the so-called
magnetars, in the context of the physics used by their mentors
(Duncan \& Thompson 1992) for proposing their occurrence in nature. 
P\'erez Mart\'{\i}nez et { al.}  (2003) argue that a fully
description of the physics taking place during the early evolution
of NSs should not overlook fundamental issues, as for
instance quantum electrodynamics effects, when discussing the
r\^ole of superstrong $B$s on the stability of
just-born NS pulsars. The {\it positive magnetization}
of the neutron matter and the {\it appearance of a ferromagnetic
configuration} in the star structure are examples of such effects.
Thus the idea of magnetars is still contentious. Despite of that
lively dispute, in this {\it Letter}, we present
theoretical arguments which alert on the potential effects of
NLED in the physics of strongly magnetized NSs.

A very interesting example of how this issue could be elusive is
provided by the recent discovery by Ibrahim et { al.} (2002), 
and its subsequent confirmation by Ibrahim et { al.} (2003), 
of cyclotron resonance features from the SGR 1806-20. It is 
well-known, in accretion-powered models, that proton ($p$) and 
$\alpha$-particles ($He$) produce, respectively, fundamental 
resonances of energy 


\be 
E_p =  ^{6.3|^{p}}_{3.2|_{He}} \left(1 + z\right)^{-1}
\left[\frac{B_{\rm Sup-Crit}}{10^{15} \rm G}\right] \; {\rm keV}
\; . 
\ee

Ibrahim et { al.} (2002;2003) showed that the 5 keV absorption 
line in the spectrum of SGR 1806-20 is 
consistent with a proton-cyclotron fundamental resonance in a
redshift-dependent super critical magnetic field ($B$) of strength:
$B_{\rm Sup-Crit} \sim 7.9 \times 10^{14} (1 + z)^{-1}$~G. This
translates into a $B_{\rm Sup-Crit} \sim 1.0 \times 10^{15}$~G,
for the mass and radius of a {\it canonical} NS ($ \rho \sim 
10^{14}$~g~cm$^{-3}$, $R \sim 10$~km, $M \sim 1.4~M_\odot$, $B 
\sim 10^{12}$~G). An estimate that agrees with the field strength 
inferred from the SGR 1806-20 spindown, i.e. from $P$ and $\dot{P}$ 
(Kouveliotou et { al.} 1998).

\section{Gravitational vs. NLED redshift}

In particular, we argue that for extremely supercritical magnetic
fields NLED effects force photons to propagate along accelerated 
curves. In case the nonlinear Lagrangean density is a function only 
of the scalar $F = F_{\mu\nu} F^{\mu\nu}$, to say an approximate 
Lagrangean 

\be
L(F) = - \frac{1}{4} F + \frac{\mu}{4} \left(F^2 + \frac{7}{4} G^2\right) 
\; , \label{E-H-LAGRANGEAN}
\ee

where $\mu = \frac{2 \alpha^2}{45} \frac{(\hbar/m c)^3}{m c^2}$, with
$\alpha = \frac{e^2}{4 \pi \hbar c}$, and $G \equiv F_{\mu\nu} F^{\ast
\; \mu\nu}$; with $F^{\ast \; \mu\nu} \equiv \frac{1}{2} \eta^{\mu\nu
\alpha \beta} F_{\alpha \beta}$, the force accelerating the photons is
given by\footnote{This Lagrangean is built up on the first two terms,
because $G = 0$ for a canonical pulsar, of the infinite series
expansion associated with the Euler-Heisenberg Lagrangean, which proved
to be valid for magnetic field strengths near the quantum
electrodynamics critical field $B\sim 10^{13.5}$~G.}

\be
k_{\alpha||\nu} k^{\nu} = 
\left(4 \frac{L_{FF}}{L_{F}} F^{\mu}_{\beta}F^{\beta\nu} k_{\mu} 
k_{\nu}\right)_{| \alpha} \; ,
\ee

where $k_{\nu}$ is the wavevector, and $L_F$ means partial derivative with
respect to $F$ (note that it does not depend on any intrinsic property
of the photons).  This feature allows for this force, acting along the
photons path, to be geometrized (Novello et { al.} 2000; Novello \&
Salim 2001) in such a way that in an effective metric: 

\be
g^{\rm eff}_{\mu\nu} = g_{\mu\nu} + g^{\rm NLED}_{\mu\nu}\;,
\ee
 
the photons follows geodesic paths, as we shall show in section (III)
in the particular case of the Lagrangean called for above. The standard
geometric procedure used in general relativity (GR) to describe the
photons can now be used upon substituting the usual metric by the
effective metric. In particular, the outcoming redshifts prove to have
now a couple of components, one due to the gravitational field and
another stemming from the $B$.

A direct insight into the GR  
$z = z(M,R)$ at the surface of a compact star could be attained from 
the identification of absorption or emission lines from it. NS mass 
($M$) can be estimated, in some cases, from the orbital dynamics of 
binary systems, while attempts to measure its radius ($R$) proceed 
via high-resolution spectroscopy (Sanwal et { al.} 2002; upon 
studying the star 1E1207.4-5209; Cottam, Paerels \& Mendez 2002; by 
analysing type-I X-ray bursts from the star EXO0748-676). In these 
systems success was achieved in determining these parameters, or the 
relation in between, by looking at excited ions near the NS surface 
(arguments favoring a strange star in EXO0748-676 are given by Xu 2003). 
Gravity effects cause the observed energies of the spectral lines of 
excited atoms to be shifted to lower values by a factor

\be
\frac{1}{(1 + z)}  \equiv \left( 1 - \frac{2 G}{c^2} \left[\frac{M}
{R }\right] \right)^{1/2} \; . \label{redshift}
\ee

Measurements of such line properties: energy, width, and polarization;  
as here called for, would lead to an indirect, but highly accurate 
estimate of the NS mass-to-radius ratio ($M/R$); and a tight constraint 
on its EoS, and to strong limits on the $B$ strength (but not 
on its configuration) at the star surface. The above analysis stands on 
whenever effects of NS $B$s are negligible. However, if the NS is 
pervaded by a super strong $B$ ($B_{\rm Sup-Crit}$), then NLED 
should be taken into account to describe the overall physics taking 
place on the pulsar surface. Our major result proves that for 
extremely high $B$s the redshift induced by NLED can be 
as high as the produced by gravity alone, thus making hard to draw any 
conclusive claim on those NS fundamental properties.

As claimed here, the shift in energy, and width, produced by the 
effective metric ``pull" of the star on laboratory known spectral 
lines, scales up directly with the strength of the effective 
potential associated to the effective metric. Thence, this shift
has two contributions: one coming from gravitational and another 
from NLED. For hyper magnetized stars, e.g. magnetars, and if the 
near surface multipole field is much stronger than the dipole 
component (see Duncan 1998 for a possible toroidal configuration in 
SGR 0526-66 based on global seismic oscillations; Section 
IV discusses implications for the cyclotron line interpretation), 
the correction factor from NLED is substantial being both 
contributions of about the same order of magnitude. Thus, there is 
the possibility, for a given field strength, for gravity effects to 
be mimicked by electromagnetic (EM) ones, and for the phenomenon to 
entangle the fixing of constraints on the $M/R$  ratio. 
This difficulty, we suggest, can be overcome by taking into account 
that the contribution of the $B$, that differs from that of 
the gravitational field which is isotropic, depends on the polarization 
$b^{\alpha}$ of the emitted photon, being different for the cases 
$B_{\alpha} b^{\alpha} = 0$ and $B_{\alpha} b^{\alpha}\neq 0$.



Our warning is, therefore, that the identification and analysis of 
spectral lines from high $B$ NSs (in outbursts) must take 
into account the two possible different polarizations of the 
received photons, in order to discriminate between redshifts 
produced either gravitationally or electromagnetially. Putting 
this result in perspective, we claim that if the characteristic $z$, 
or $M/R$ ratio, were to be inferred from this type of sources care 
should be taken since for this superstrong $B$ such $z$ becomes of 
the order of the gravitational one expected from a canonical NS. It is, 
therefore, not clear whether one can cathegorically assert something 
about the, e.g. SGR 1806-20, $M/R$ ratio under such dynamical conditions. 
We prove this claim next.


\section{The Model}


The propagation of photons in NLED has been examined by several
authors (Bialynicka-Birula \& Bialynicki-Birula 1970; Garcia \&
Plebanski 1989; Dittrich \& Gies 1998; De Lorenci, Klippert, Novello \&
Salim 2000). In the case of geometric optics, where the photon
propagation can be identified with the propagation of discontinuities
of the EM field in a nonlinear regime, a remarkable property appears:
the discontinuities propagate along null geodesics of an effective
geometry which depends on the EM field of the background (Novello et
{ al.} 2000; Novello \& Salim 2001).  According to quantum
electrodynamics,  in Heisenberg \& Euler (1936) approximation (see
also Schwinger 1951), a vacuum has nonlinear properties, and these
novel properties of photon propagation in NLED can show up, in
principle, in photons propagating in a vacuum. In this specific case, the
equations for the EM field in a vacuum coincide in their form with the
equations of continua in which the electric and magnetic permittivity
tensors $\epsilon_{\alpha\beta}$ and $ \mu_{\alpha\beta}$ are functions
of the electric and magnetic fields, determined by some observer
represented by its velocity 4-vector $V^{\mu}$ (Denisov, Denisova
\& Svertilov 2001a, 2001b; Denisov \& Svertilov 2003). We should
remark that this first order approximation is valid for magnetic fields
smaller than $ B_{q} $, a parameter that will be defined bellow.  In
curved spacetime, these equations are written as

\begin{equation}
D^{\alpha}_{||\alpha}=0, \hskip 0.5 truecm B^{\alpha}_{||\alpha}=0\; ,
\end{equation}


\begin{equation}
D^{\alpha}_{||\beta}\frac{V^{\beta}}{c} +
\eta^{\alpha\beta\rho\sigma}V_{\rho}H_{\sigma||\beta}=0,
\end{equation}

\begin{equation}
B^{\alpha}_{||\beta}\frac{V^{\beta}}{c} -
\eta^{\alpha\beta\rho\sigma}V_{\rho}E_{\sigma||\beta}=0,
\end{equation}

where the double vertical bars ``$||$'' stand for covariant derivative, and $\eta^{\alpha\beta\rho\sigma}$ is the completely antisymmetric Levi-Civita tensor. The 4-vectors representing the EM field are defined as usual in 
terms of the EM field tensor  $F_{\mu\nu}$ and polarization tensor 
$P_{\mu\nu}$

\begin{equation}
E_{\mu}=F_{\mu\nu}\frac{V^{\nu}}{c}, \hskip 0.5 truecm B_{\mu} = 
F^{*}_{\mu\nu}\frac{V^{\nu}}{c}\; ,
\end{equation}

\begin{equation}
D_{\mu}=P_{\mu\nu}\frac{V^{\nu}}{c}, \hskip 0.5 truecm H_{\mu} = 
P^{*}_{\mu\nu}\frac{V^{\nu}}{c}\; ,
\end{equation}

where the dual tensor $X^{*}_{\mu\nu}$ is defined as $ X^{*}_{\mu\nu} = \frac{1}{2}\eta_{\mu\nu\alpha\beta} X^{\alpha\beta}\;$,
for any antisymmetric second-order tensor $X_{\alpha\beta}$.
The meaning of the vectors $D^{\mu}$ and $H^{\mu}$ comes from the
Lagrangean of the EM field,  and in the case of a vacuum they are   

\begin{equation}
H_{\mu}= \mu_{\mu\nu}B^{\nu},  \hskip 0.5 truecm D_{\mu}=\epsilon_{\mu\nu}E^{\nu}\; ,
\end{equation}

where the permittivity tensors are given as

\begin{equation}
\mu_{\mu\nu}=\left[1+ \frac{2 \alpha}{45 \pi B^2_q}\left(B^2-E^2\right)
\right]h_{\mu\nu} - \frac{7 \alpha}{45 \pi B^2_q} E_{\mu} E_{\nu}\; ,
\label{permittivity-mu}
\end{equation}

\begin{equation}
\epsilon_{\mu\nu}=\left[1+ \frac{2 \alpha}{45 \pi B^2_q}\left(B^2-E^2\right)
\right]h_{\mu\nu} + \frac{7 \alpha}{45 \pi B^2_q} B_{\mu} B_{\nu}\; .
\label{permittivity-epsilon}
\end{equation}

In these expressions $\alpha$ is the EM coupling constant
$(\alpha=\frac{e^2}{hc} = \frac{1}{137})$ and $B_q$ is a quantum
electrodynamic parameter,  $B_q = \frac{m^2c^3}{eh} =
4.41\times10^{13}$~G, also known as the Schwinger critical $B$-field, 
i.e., $B_q \equiv B_{crit}$.  The tensor $h_{\mu\nu}$ is the
metric induced in the reference frame perpendicular to the observers, 
determined by the vector field $V^\mu$. Our main concern in this paper
is the behavior of NLED in either a pulsar or a 
magnetar, so in this particular case,  $E^\alpha = 0$, 
$\epsilon^\alpha_\beta = \epsilon h^\alpha_\beta + \frac{7 \alpha} {45
\pi B^2_q} B^\alpha B_\beta $ and $\mu_{\alpha\beta} = \mu
h_{\alpha\beta}$. The scalars $\epsilon$ and $\mu$ can be read directly
from Eqs.(\ref{permittivity-mu},\ref{permittivity-epsilon}) as $
\epsilon \equiv \mu = 1 + \frac{2 \alpha}{45 \pi B^2_q}(B^2)$.  We
will deal with light propagation in NLED in
optical approximation. The EM wave is represented by 3-surfaces of
discontinuities of the EM field that propagate in the nonlinear
background. As we show below, the EM wave propagation can be described as
if the metric of the background were changed from its original form
determined by GR into another effective metric that depends on
the dynamics of the background EM field. This formalism allows us to
use the well-known results from Riemann geometry largely applied in
GR.

Following Hadamard (1903), the surface of discontinuity of the
EM field is denoted by $\Sigma$. The field is continuous
when crossing $\Sigma$, while its first derivative presents a finite
discontinuity, specified as follows

\begin{equation}
[B^\mu]_{\Sigma} = 0, \hskip 0.3 truecm \left[\partial_\alpha B^\mu\right]_{\Sigma} 
= b^\mu k_\alpha, \hskip 0.3 truecm \left[\partial_\alpha E^\mu 
\right]_{\Sigma} = e^\mu k_\alpha \protect\label{eq14}\; ,
\end{equation}


where the symbol 

\begin{equation}
\left[J\right]_{\Sigma}=\lim (J_{\Sigma + \delta}-J_{\Sigma
- \delta}) \label{discontinuity} \; 
\end{equation}

represents the discontinuity of the arbitrary function $J$ through the
surface $\Sigma$. The tensor $f_{\mu\nu}$ is called the discontinuity
of the field,  and $k_{\lambda} = \partial_{\lambda} \Sigma $ is the
propagation vector. Applying conditions (\ref{eq14}) and
(\ref{discontinuity}) to the field equations in the particular case of
$E^\alpha = 0$, we obtain the constrains $e^{\mu} \epsilon_{\mu\nu}
k^{\nu} = 0$ and $b^{\mu} k_{\mu} = 0$ and the following equations for
the discontinuity fields $e_\alpha$ and $b_\alpha$:

 \begin{equation}
 \epsilon^{\lambda\gamma}e_{\gamma}k_{\alpha} \frac{V^\alpha}{c} + \eta^{\lambda\mu\rho\nu} \frac{V_\rho}{c} \left(\mu b_\nu k_\mu - 
{\mu^\prime} \lambda_\alpha B_\nu k_\mu \right)=0 \; , \label{dispersao-1}
 \end{equation}

 \begin{equation}
 b^{\lambda}k_{\alpha}\frac{V^\alpha}{c} - \eta^{\lambda\mu\rho\nu} \frac{V_\rho}{c} \left(e_\nu  k_\mu \right)=0 \; .  \label{dispersao-2}
 \end{equation}

Isolating the discontinuity field from (\ref{dispersao-1}), substituting
in equation (\ref{dispersao-2}), and expressing the products of the
completely anti-symmetric tensors $\eta_{\nu\xi\gamma\beta} \eta^{\lambda\alpha\rho\mu}$ in terms of delta functions, we obtain

\begin{eqnarray}
b^{\lambda}(k_\alpha k^\alpha)^2 + \left(\frac{\mu'}{\mu}l_\beta b^\beta k_\alpha B^\alpha + \frac{\beta B_\beta b^\beta B_\alpha k^\alpha} {\mu-\beta B^2}\right) k^{\lambda} \nonumber  \\ 
+ \left(\frac{\mu'}{\mu l_\alpha b^\alpha} \left[(k_\beta V^\beta)^2 
(k_\alpha k^\alpha)^2\right] - \frac{\beta B_\alpha b^\alpha (k_\beta k^\beta)^2}{\mu-\beta B^2}\right) B^{\lambda} \nonumber \\ 
- \left(\frac{\mu'}{\mu} l_\mu b^\mu k_\alpha  B^\alpha k_\beta V^\beta\right)V^{\lambda} = 0 \; . \label{dispersion-01} 
\end{eqnarray}

This expression is already squared in $k_\mu$ but still has an unknown
$b_\alpha$ term. To get rid of it, one multiplies by $B_\lambda$, to
take advantage of the EM wave polarization dependence.  By noting that
if $B^\alpha b_\alpha = 0$ one obtains the {\it dispersion relation} by
separating out the $k^{\mu} k^{\nu}$ term, what remains is the (-)
effective metric. Similarly, if $B_\alpha b^\alpha \neq 0$, one simply
divides by $B_\gamma b^\gamma$ so that by factoring $k^{\mu} k^{\nu}$, 
what results is the (+) effective metric. For the case $ B_\alpha
b^\alpha = 0$, one obtains

 \begin{equation}
  g^{\alpha\beta} k_\alpha k_\beta = 0 \; .
 \end{equation}
 
 whereas for the case $ B_\alpha b^\alpha \neq 0$, the result is

 \begin{eqnarray}
 \left[\left(1+\frac{\mu'B}{\mu} + \frac{\beta B^2}{\mu-\beta B^2}\right)
  g^{\alpha\beta} - \frac{\mu'B}{\mu}\frac{V^\alpha V^\beta}
  {c^2} + \left(\frac{\mu'B}{\mu} + \nonumber \right. \right. \\
 + \left. \left. \frac{\beta B^2}{\mu-\beta B^2}\right) l^{\alpha}l^{\beta}\right] k_{\alpha} k_{\beta} = 0 \; ,
  \end{eqnarray}

 where by $( ' )$ we mean $\frac{d}{dB}$, and we define $\beta = \frac{7 \alpha}{45 \pi B^2_q}$, and $l^\mu \equiv \frac{B^\mu}{|B^\gamma B_\gamma|^{1/2}}$.

 From the above expressions we can read the effective 
metric $g^{\alpha\beta}_{+}$ and $g^{\alpha\beta}_{-}$, where the 
labels ``+'' and ``-'' refers to extraordinary and ordinary polarized 
rays, respectively. To determine the redshift we need the covariant 
form of the metric tensor, obtained from the expression 
$ g_{\mu\nu} g^{\nu\alpha} = \delta^\alpha_{\; \; \; \mu } \;$. It reads

\begin{equation}
g^{-}_{\mu\nu} = g_{\mu\nu} \; \label{g-}
 \end{equation}

and

  \begin{eqnarray}
g^{+}_{\mu\nu} & = & \left(1+\frac{\mu'B}{\mu} + \frac{\beta
B^2}{\mu-\beta B^2}\right)^{-1} g_{\mu\nu} \nonumber \\
& + & \left[\frac{\mu'B}{\mu (1+\frac{\mu'B}{\mu}+ \frac{\beta
B^2}{\mu-\beta B^2})(1+ \frac{\beta B^2}{\mu-\beta B^2})}
\right] \frac{V_{\mu}V_{\nu}}{c^2} \nonumber \\
& + & \left(\frac{\frac{\mu'B}{\mu}+ \frac{\beta B^2}{\mu-\beta
B^2}}{1+\frac{\mu'B}{\mu} + \frac{\beta B^2}{\mu-\beta B^2}} \right) 
\; \; l_\mu l_\nu \; . \label{l-dependent}
  \end{eqnarray}
 
The function $\frac{\mu'B}{\mu}$ can be expressed in terms of the 
magnetic permissivity of the vacuum, and is given as

  \begin{equation}
  \frac{\mu'B}{\mu} = 2\left(1-\frac{1}{\mu}\right)\; .
  \end{equation}

In the particular case that we are focusing on, both the emitter and 
observer are in inertial frames, that is, $V^\mu = {\delta^\mu_0}/
(g_{00})^{1/2} $; therefore, the components of both effective metrics 
above become coincident and given as

\begin{equation}
g^{\rm eff}_{00} = \frac{g_{00}}{ 1 + \frac{\beta B^2}{\mu - \beta B^2} }\; .
 \end{equation}

The general expression for the   redshift is then given as

  \begin{equation}
\frac{\nu_{B}}{\nu_{A}} = \frac{\lambda_{A}}{\lambda_{B}} = 
\left[\frac{g_{00}(e)}{ g_{00}(o)}\right]^{\frac{1}{2}} \; ,
  \end{equation}

  or

  \begin{equation}
z = \frac{\lambda_{B}-\lambda_{A}}{\lambda_{A}} =  
\left[\frac{g_{00}(e)}{g_{00}(o)}\right]^{-\frac{1}{2}} - 1.
  \end{equation}

where $g_{00}(e)$ and $ g_{00}(o)$ stand for the time-time effective 
metric components at emission and observation, respectively. Hence, 
for observations very far from the star the redshift can be approximated
as 

 \begin{equation}
 z + 1 = \left[ \frac{g_{00}(o)}{g_{00}(e)} \right]^{\frac{1}{2}} =
\frac{ \left(1 - \frac{2GM} {c^2 R} \right)^{-\frac{1}{2}} }{ \left[ 1 + 
\frac{ \beta B^2}{\mu - \beta B^2} \right]^{\frac{1}{2}} }
\label{redshift-B}\; .
 \end{equation}

\begin{equation}
 z + 1 \simeq \frac{ \left(1 - \frac{2 G M} {c^2 R} \right)^{-\frac{1}{2} } }
{ \left[ 1 + \beta B^2 \right]^{\frac{1}{2}} } = \frac{\left(1-0.3\frac{M}{R} \right)^{-\frac{1}{2}} } {\left[1 + 0.19 B^2_{15} \right]^{\frac{1}{2}}} 
\label{redshift-C}\; 
 \end{equation}


where $M$ is the star mass in units of $M_\odot$,  $R$ its radius in units of 10~km, and $B_{15}$ is the B-field in units of $10^{15}$~G.

Note, however, that in the present case the correction on the
gravitational redshift brought to $z$ by the nonlinear contribution of
the magnetic field does depend on the polarization $b^\mu$ of the
emitted photons. Therefore, it is straightforward to verify that
because of the appearance of the two different effective metrics in
equations (\ref{l-dependent},\ref{g-}), which exihibit the phenomenon
of birefringence, one may {\it in principle} disentangle the two
components of the total pulsar surface redshift by a direct
observation.

\section{Discussion and Conclusion}

The 5.0 keV feature discovered with the {\it Rossi X-ray Timing Explorer} 
is strong, with an
equivalent width of $\sim 500$~eV and a narrow width of less than
0.4~eV (Ibrahim et { al.} 2002, 2003). When these features are viewed
in the context of accretion models, one arrives to a $M/R > 0.3$
M$_\odot$ km$^{-1}$, which is inconsistent with NSs, or requires a
low $B \sim (5-7) \times 10^{11}$~G, which is said not to correspond to
any SGRs (Ibrahim et { al.} 2003). In the magnetar scenario,
meanwhile, the features are plausibly explained as being ion-cyclotron
resonances in an ultra strong $B$-field, $B_{\rm sc} \sim 10^{15}$~G,
whose energy and width are close to model predictions (Ibrahim et {
al.} 2003).  According to Ibrahim et { al.} (2003), the confirmation
of this findings would allow to estimate the gravitational redshift,
mass, and radius of the supposed magnetar SGR 1806-20.

Here we point out the possibility that this feature could also be due
to NLED in the same superstrong $B$-field of SGR 1806-20, as suggested
by equation (\ref{redshift-B}). To obtain our conclusion, we used $B
\sim 5 \times 10^{15}$~G, which is within the uncertainty of the
$B$-field strength estimate from $P$ and $\dot{P}$ and the likely
$B$-field near-surface multipole structure, as suggested by various
authors in the field. In particular, Duncan (1998) interpreted the
23~ms {\it global} oscillations observed in the ``magnetar-like" object
SGR 0526-66 as being a fundamental {\it toroidal mode}, assuming a
field $B \sim 4 \times 10^{15}$~G lies underneath the star crust. Other
authors hintat the coexistence of {\it poloidal} configurations as
well. For such fields, the cyclotron viewpoint could be sustained only
whenever the dipole component is the dominant emission mechanism. If
this were the case, no conclusive assertion about the $M/R$ ratio of
the compact star glowing in SGR 1806-20 could be consistently made,
since the NLEDredshift  might well be mimicking the standard
gravitational redshift associated with the pulsar surface. More
fundamentally yet, if new spectral lines were measured with high
precision (as in Cottam et al. 2002) from heavy elements in a compact
object with fields $B\gtrsim 10^{15}$~G, then the $\Delta z \gtrsim
10$~ z-correction brought by NLED would prove critical regarding
both its $M/R$ ratio and its EoS.

As a worthy remark, the attentive reader must realize that there exists
a hidden divergence in the effective metric here derived. It appears
when the magnetic field strength achieves values around $B
\longrightarrow B_q \sim 10^{13.5}$~G. We stress that such a divergence
is inherent to the sort of approximation we are using for, that is, the
Heisenberg \& Euler (1936) Lagrangean, which is not an exact one, of
which we just take into account only the first term in its expansion.
We advance, meanwhile, that such divergence can be removed by taken
advantage of a very different sort of nonlinear electrodynamics 
Lagrangean, as the exact one introduced by Born \& Infeld (1934). 
This new approach is matter of a forthcoming communication.

\acknowledgements 

{ HJMC thanks Prof. J. A. de Freitas Pacheco for fruitful discussions 
and Observatoire de la C\^ote d'Azur, NICE, for hospitality. 
Support from Funda\c c$\tilde{a}$o de  Amparo \`a Pesquisa do Estado 
de Rio de Janeiro (FAPERJ/Brazil) through the Grant-in-Aid 151.684/2002 is 
acknowledged. J.M.S. acknowledges Conselho Nacional de Desenvolvimento 
Cient\'{\i}fico e Tecnol\'ogico (CNPq/Brazil) for the Grant No. 302334/2002-5.}


\end{document}